\documentclass[pss,fleqn]{w-art}
\usepackage{times}
\usepackage{w-thm}
\usepackage[english,french]{babel}
\usepackage[]{graphicx}
\begin{document}
\DOIsuffix{theDOIsuffix}
\pagespan{3}{}



\title{Comparison between models of insulator and semiconductor thin films islanding}


\author[Sh. First Author]{Fran\c cois Lallet\footnote{Corresponding
     author: e-mail: {\sf f\_lallet@ensci.fr}, Phone: +33\,555\,452\,222,
     Fax: +33\,555\,790\,998}\inst{1}} \address[\inst{1}]{SPCTS, UMR-CNRS 6638 47 \`a 73 Avenue Albert Thomas, 87065 Limoges cedex, FRANCE}
\author[Sh. Second Author]{Alain Dauger\footnote{Second author: e-mail: {\sf a\_dauger@ensci.fr}, Phone: +33\,555\,452\,224,
     Fax: +33\,555\,790\,998}\inst{1}}
\author[Sh. Third Author]{Nathalie Olivi-Tran\footnote{Third author: e-mail: {\sf n\_olivi-tran@ensci.fr}, Phone: +33\,555\,452\,247, Fax: +33\,555\,790\,998}\inst{1}}
\begin{abstract}
The synthesis of self-organized quantum dots (QD's) can be achieved through bottom up layer by layer deposition processes as chemical vapor deposition (CVD) or physical vapor deposition (PVD). However, QD's may also be synthesized via sol-gel route, which involves a spontaneous evolution from thin films to discrete QD's without further deposition.
The aim of the paper is to discuss and compare the physical phenomena involved in QD's formation which initiate from thin film surface roughening between PVD-CVD and sol-gel synthesis approaches.
We propose two simple physical models which are relevant to explain the fundamental differences between those methods.
\end{abstract}
\maketitle                   





\section{Introduction}

Many authors have presented and studied epitaxial QD's growth through PVD-CVD synthesis processes. The physical phenomena associated with such approaches have been widely studied both theoretically \cite{Nurminen}\cite{Kalke}\cite{Liu} and experimentally \cite{Alchalabi}\cite{Nie}, in particular through the model system Ge/Si \cite{Sutter}\cite{Tromp}\cite{Capellini}\cite{Portavoce}\cite{Chen}\cite{Wagner1}\cite{Wagner2}, because of its promising technological applications. However the synthesis of QD's epitaxially grown on a crystalline substrate can be achieved through a sol-gel approach. Indeed, Bachelet et al \cite{Bachelet} have recently synthesized and studied the microstructure of zirconia QD's grown on a c-cut sapphire substrate during thermal annealing of a zirconia precursor xerogel thin film deposited on the substrate by sol-gel dip-coating.

The aim of this article is to discuss the physical phenomena involved in these processes through the analysis of two physical models, based on energetic considerations, which have been developed for numerical simulations. We suppose, for the sake of simplicity, that QD's synthesis is achieved without nucleation.

\section{Physical models}

We have developed a Monte Carlo (MC) algorithm to simulate the islanding of a thin film on a perfect crystalline subtrate during thermal annealing without further deposition \cite{Lallet}. The numerical thin film is divided into mesoscopic domains which are characterized by their height and their crystallographic orientation with regard to their nearest neighbors ($NN$) and the substrate. At each MC step, a domain $i$ is chosen at random and the probability $P_i$ to change its height ($h_i$) and/or its crystallographic orientation ($c_i$, $d_i$) is calculated through the classical Metropolis scheme \cite{Metropolis}. The energy of the domain $i$ with regard to its nearest neighbors is expressed as: 
\begin{equation}
E_i= \gamma_1 \left(\frac{\ell^2}{h_i}+ \ell\right) \sum_{j=1}^{NN}  (c_i-c_j) + \\
\gamma_2 \left(\frac{\ell^2}{h_i}+\ell\right) \sum_{j=1}^{NN} (d_i-d_j)+ \\
Y (1+\nu) \sqrt{\frac{ D_s \gamma_s \Delta t}{k_B T}} \ell^2 \sum_{j=1}^{NN}(h_i-h_j)
\end{equation}
where the first and second term of the right hand side of the equality correspond to the interfacial energy of the domain with regard to the substrate and to its neighbors respectively. The third term correspond to the surface energy related to the heights of the domains. $\gamma_1$ is the boundary surface tension (domain/domain), $\gamma_2$ the interfacial surface tension (domain/substrate), $\ell$ the distance between domain $i$ and its nearest neighbors,  $Y$ the Young modulus, $\nu$ the Poisson's ratio, $D_s$ the surface diffusion coefficient, $\gamma_s$ the free surface tension, $k_B$ the Boltzmann constant and T the absolute temperature.

We present another physical model which allows one to discuss, at least qualitatively, the main parameters responsible for the morphological evolution of a thin film epitaxially grown on a perfect crystalline substrate through a deposition process (PVD or CVD). This model is inspired from the previous works of Kawamura \cite{Kawamura} and Russo \cite{Russo} who established MC algorithms at the atomic scale to model QD's formation modes during deposition processes. The energy of an adatom $i$ is computed as the sum of its bonding energy and elastic energy \cite{Kawamura}\cite{Russo}:
\begin{equation}
E_i=NE_B - (E_\textrm{with adatom i}-E_\textrm{without adatom i})
\end{equation}
where $N$ is the number of chemical bonds of the adatom $i$ with its neighbors, $E_B$ is the energy of a chemical bond and $E$ the total elastic energy. From equation (2), we deduce a model in which an epitaxial monocrystalline thin film  is deposited on a perfect crystalline substrate. We model this film as a cubic array of mesoscopic domains $i$ of height $h_i$ and width $\ell$. Each domain is submitted both to its surface tension $\gamma_s$ and to an elastic stress field induced by the lattice mismatch between the film and the substrate, $\epsilon$. The volume energy density of a thin film of initial thickness $h$, due to free surface energy, is expressed as $\gamma_s/h$. Thus, the surface tension of a domain $i$ can be written as $(\gamma_sh_i)/h$ and therefore the surface energy of domain $i$ (related to $NE_B$) induced by the free surface energy is $(\ell^2\gamma_sh_i)/h$ where $\ell^2$ is the free surface area at the top of each domain. The elastic stress field leads to an elastic energy inside each domain of the film. This elastic energy (related to $E_\textrm{with adatom i}-E_\textrm{without adatom i}$) can be expressed through linear elasticity theory. We suppose for the sake of simplicity that the elastic stress tensor is diagonal (for example for cubic phased materials). 
With the same assumption, $\epsilon$, which is the lattice mismatch between the film and the substrate, is constant in the horizontal plane. Thus the resulting force in the horizontal plane reduces to $2Y\epsilon \ell h_i$ with $\ell h_i$ the area of surfaces of domain $i$ perpendicular to the $x$ and $y$ axis. Futhermore, the resulting force supported by the domain $i$ on the vertical axis $z$ is equal to $\nu Y\epsilon \ell^2$. Therefore, the resulting elastic energy related to domain $i$ and induced by the lattice mismatch $\epsilon$ is given by the work of this elastic force, for a characteristic displacement $\epsilon \ell$  in the horizontal plane: 
\begin{equation}
E=Y\epsilon(2\ell h_i+\nu \ell^2)\epsilon \ell
\end{equation}
Consequently, the energy of a domain $i$ with regard to its neighbors $j$ might be written:
\begin{equation}
E_i=\ell^2\left(\frac{\gamma_s}{h}- 2Y\epsilon^2\right)\sum_{j=1}^{NN}(h_i-h_j)
\end{equation}

\section{Discussion}

The formation of QD's without nucleation is achieved through the evolution of the roughness of a thin film until nanometer scale islands are clearly identified. In the numerical models, the evolution of the roughness of a film is simulated by the evolution of the heights of the discrete mesoscopic domains composing the virtual thin film.
The energy of a mesoscopic domain $i$ of a thin film synthesized via sol-gel route is calculated with equation (1). One can see that the roughness of the surface of the film is driven by surface diffusion and surface tension. The energy of a mesoscopic domain $i$ of a thin film synthesized via a deposition process is calculated through equation (4). This equation allows one to understand that the roughness of the film is driven by a competition effect between surface and stress energies.

The probability of changing the height of a domain with regard to equation (1) follows a monotonic tendency
with regard to the intrinsic parameters of the thin film. Figure (1) presents the evolution of the islanding of numerical thin films deposited on a substrate by sol-gel dip-coating after $10^7$ MCS as a function of the initial thickness of the film. The annealing temperature is fixed, so all the parameters of the equation are fixed. One can see that as the initial thickness of the film increases, the substrate is less dewetted. Indeed, as the initial thickness increases ($h=4nm$), the amplitude of the roughness which appears at the top of the layer can not reach the substrate as fast as for thiner films ($h=1nm$). 

The probability of changing the height of a domain with regard to equation (4) depends on the relative values of the parameters $\gamma_s/h$ and $2Y\epsilon^2$. Indeed, if $\gamma_s/h>2Y\epsilon^2$ then the energy of the domain $i$ is expressed as $E_i=K\sum_{j=1}^{NN}(h_i-h_j)$ (where $K=\gamma_s/h- 2Y\epsilon^2$) with $K>0$, whereas $K<0$ otherwise. On the one hand, if the lattice mismatch $\epsilon$ is low and/or with a low film thickness ($K>0$), then the effect of surface energy is predominant. Therefore the probability of roughening is low and the film remains almost flat during the growing of the film which is described through the Frank-van-der-Merke (F-M) deposition process (or ALD Atomic Layer Deposition process). On the other hand, if the numerical values of $\epsilon$ and/or film thickness are high ($K<0$), then the probability of roughening increases; therefore the effect of stress field inside the film is predominant which leads to the formation of islands. The surface roughening stage can either initiate from the top of a wetting layer, which is called  the Stranski-Krastanov (S-K) growth mode, or directly from the top of the substrate without previous deposition of a wetting layer ($\epsilon$ very high) which is the Volmer-Weber (V-W) growth mode. 
Figure (2) illustrates the numerical calculations of the islanding process through the effect of the initial film thickness (top, $\epsilon$=0.04) while the bottom presents this evolution with regard to lattice mismatch (bottom, $h$=2nm) after $10^7$ MCS. We focus on the transition of $K$ from positive to negative values as the islands are formed when $K<0$. One can note that the increase of $h$ or $\epsilon$ leads to a higher number of QD's, which is contrary to the evolution of a thin film synthesized via sol-gel processing where the higher thickness of the film leads to a lower number of QD's. Moreover, it is clear that QD's formation is more sensitive to the variation of the lattice mismatch than to the variation of the initial thickness of the film as $E_i$ is a function of $\epsilon^2$ and $1/h$. Thus the model describes qualitatively the evolution from a F-M or S-K growth mode ($K>0$ and $K\approx0$ respectively) to a V-W growth mode $K<0$ as the number of QD's synthesized grows with the increase of $\epsilon$ and/or $h$. Those tendencies have already been reported for the Ge/Si system \cite{Voigtlander}.

\section{Conclusion}
Our simple energetic models allow us to describe the main tendency of QD's formation from thin film synthesized either via deposition processes (PVD-CVD) or sol-gel dip-coating.

The main difference between those approaches is that thin films synthesized via sol-gel route lead to a greater number of QD's when they are thin while thin films synthesized via deposition processes demonstrate the opposite behaviour. 
This result is explained through equations (1) and (4) which point out that surface roughening of thin films synthesized via sol-gel route is driven by surface diffusion and surface tension, whereas it is the consequence of a competitive mechanism between elastic stress energy and surface tension for thin films synthesized layer by layer in PVD or CVD processes.

\begin{figure}
\begin{center}
\includegraphics[scale=0.5]{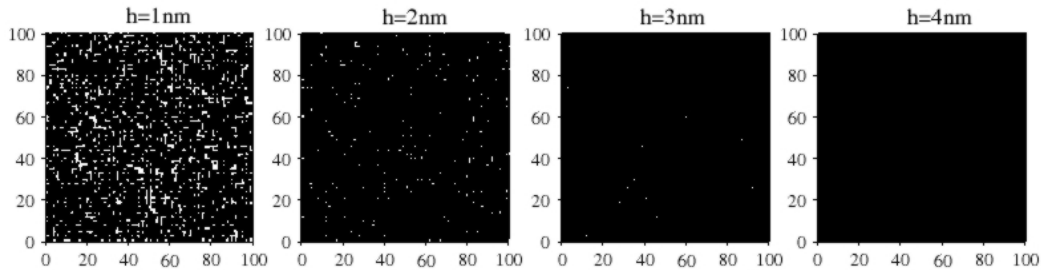}
\caption{Simulation of the islanding process of numerical thin films without deposition (sol-gel) as a function of the initial thickness of the film after $10^7$ MCS.}
\end{center}
\end{figure} 

\begin{figure}
\begin{center}
\includegraphics[scale=0.5]{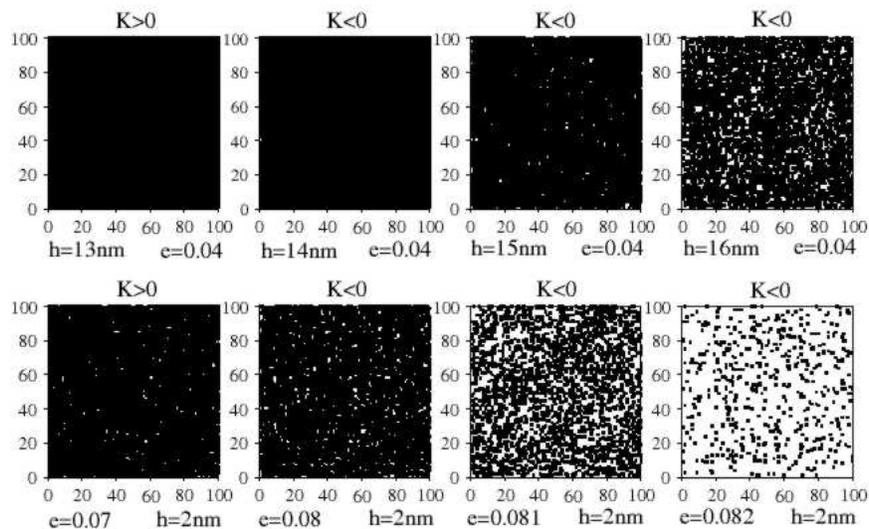}
\caption{Simulation of the islanding process of numerical thin films with deposition (PVD-CVD) as a function of the initial thickness of the film (top) and of the lattice mismatch (bottom) after $10^7$ MCS.}
\end{center}
\end{figure}


\begin{thebibliography}{18}
\bibitem{Nurminen} L. Nurminen, A. Kuronen and K. Kaski Phys. Rev. B \textbf{63}, 035407 (2000).
\bibitem{Kalke} M. Kalke, D.V. Baxter, Surf. Sci. {\bf 477}, (2001) 95-101.
\bibitem{Liu} P. Liu, Y.W. Zhang, C. Lu Phys. Rev. B {\bf 68}, 035402 (2003).
\bibitem{Alchalabi} K. Alchalabi, D. Zimin, G. Kostorz, H. Zogg Phys. Rev. Lett. {\bf 90} 026104 (2003).
\bibitem{Nie} J.C. Nie, H. Yamasaki, Y. Mawatari Phys. Rev. B {\bf 70}, 195421 (2004).
\bibitem{Sutter} P. Sutter, M.C. Lagally Phys. Rev. Lett. {\bf 84}, 4637 (2000).
\bibitem{Tromp} R. M. Tromp, F.M. Ross and M.C. Reuter Phys. Rev. Lett. {\bf 84}, 4641 (2000).
\bibitem{Capellini} G. Capellini, M. De Seta, F. Evangelisti Mater. Sci. Eng. B {\bf 89}, (2001) 184-187.
\bibitem{Portavoce} A. Portavoce, M. Kammler and  R. Hull Phys. Rev. B {\bf 70}, 195306 (2004).
\bibitem{Chen} P.S. Chen, Z. Pei, Y.H. Peng, S.W. Lee, M.-J. Tsai  Mater. Sci. Eng. B {\bf 108}, (2004) 213-218.
\bibitem{Wagner1} R.J. Wagner and E. Gulari Phys. Rev. B {\bf 69}, 195312 (2004).
\bibitem{Wagner2} R.J. Wagner, E. Gulari Surf. Sci. {\bf 590}, (2005) 1-8.
\bibitem{Bachelet} R. Bachelet, A. Boulle, B. Soulestin, F. Rossignol, R. Guinebreti\`ere and A. Dauger, submitted to Thin Solid Films.
\bibitem{Lallet} F. Lallet, R. Bachelet, A. Dauger and N. Olivi-Tran ArXiv: cond-mat/0512228.
\bibitem{Metropolis} N. Metropolis, A.W.  Rosenbluth, M.N. Rosenbluth, A.T. Teller, E.J. Teller, Chem. Phys. {\bf 21} (1953) 1087.
\bibitem{Kawamura} T. Kawamura, T. Natori, Surf. Sci. {\bf 438}, (1999) 148-154.
\bibitem{Russo} G. Russo, P. Smereka, J. Comp. Phy. {\bf 214}, (2006) 809-828.
\bibitem{Voigtlander} B. Voigtl\"ander, Surf. Sci. Rep. {\bf 43}, (2001) 127-254.
\end{thebibliography}
\end{document}